\def\bey{\begin{eqnarray}}
\def\eey{\end{eqnarray}}
\def\be{\begin{equation}}
\def\ee{\end{equation}}
\def\ba{\begin{array}}
\def\ea{\end{array}}

\def\om{\omega}

\def\pp{\partial}
\def\pp{\partial}

\documentclass{elsart}
\usepackage{amsmath,bm}
\usepackage{graphicx}
\usepackage{dcolumn}
\usepackage[colorlinks,linkcolor=blue,citecolor=blue,urlcolor=blue,anchorcolor=blue]{hyperref}

\usepackage[normalem]{ulem}  
\usepackage[dvips]{color} 

\begin{document}

\begin{frontmatter}
\title{ Liquid-gas phase transition in hot asymmetric nuclear matter with density-dependent relativistic mean-field
models }
\author{Guang-Hua Zhang$^1$ and Wei-Zhou Jiang$^{1,2}$}
\address{  $^1$ Department of Physics, Southeast University,
Nanjing 211189, China}\address{ $^2$ National  Laboratory of Heavy Ion
Accelerator, Lanzhou 730000,China}
\date{}

\begin{abstract}
\baselineskip18pt The liquid-gas phase transition in hot asymmetric
nuclear matter is studied  within density-dependent relativistic
mean-field models where the density dependence is introduced
according to the Brown-Rho scaling and constrained by available data
at low densities and empirical properties of nuclear matter. The
critical temperature of the liquid-gas phase transition is obtained
to be 15.7 MeV in symmetric nuclear matter falling on the lower edge
of the small experimental error bars. In hot asymmetric matter, the
boundary of the phase-coexistence region is found to be sensitive to
the density dependence  of the symmetry energy. The critical pressure
and the area of phase-coexistence region increases clearly with the
softening of the symmetry energy. The critical temperature of hot
asymmetric matter separating the single-phase region from the
two-phase region  is analyzed to have a moderate sensitivity to the
symmetry energy and is higher for the model possessing the softer
symmetry energy.

\end{abstract}


\end{frontmatter}

\section{Introduction}
The determination of the properties of hadronic matter at finite
temperature and density is a fundamental problem in   nuclear
physics. At low densities, the so-called liquid-gas (LG) phase
transition in nuclear matter may  occur at sufficiently low
temperatures due to the van der Waals behavior of the nucleon-nucleon
interaction. The study of the LG phase transition in intermediate
energy heavy-ion collisions is of considerable interest over the last
three decades. In the past, properties of the nuclear LG phase
transition have been explored both experimentally and theoretically
in a variety of works
~\cite{dql78,gfb83,hj84,hqs91,hb95,ygm97,qwl02,mb99,qwl00,jbn02,pmj04,ri10,cw11,ag12,jx07,bk09,bal08}.
The calculated critical temperature of symmetric nuclear matter lies
in a wide range, e.g., 13-24 MeV, for various phenomenological
models~\cite{mb99,qwl00,jbn02,pmj04,ri10,cw11}.

To understand better the features of the LG phase transition in hot
asymmetric nuclear matter, it is imperative to employ the nuclear
equation of state (EOS) of asymmetric matter. Recent progress in
experiments with radioactive beams provides us a great opportunity to
constrain the nuclear EOS of asymmetric matter. However, the nuclear
EOS of asymmetric matter, especially the density dependence of the
symmetry energy, is still rather poorly
known~\cite{bal08,bab00,cjh01}. The study of the LG phase transition
may open a possible window to constrain the nuclear EOS of asymmetric
matter.  In the past, the sensitivity of the LG phase transition to
the density dependence of the symmetry energy was explored with the
non-relativistic models in Ref.~\cite{jx07}. Similar work with the
nonlinear relativistic mean-field (RMF) models can be found in
Ref.~\cite{bk09}. In this work, we study the thermodynamic properties
of the LG phase transition in hot asymmetric nuclear matter with
density-dependent RMF models.

The density-dependent RMF models adopted in present
work~\cite{ji07a,ji07b} feature the chiral limits at high densities
with the in-medium hadron properties according to the Brown-Rho (BR)
scaling~\cite{brown91,brown07}. These models achieved satisfactory
success in describing the nuclear EOS of asymmetric matter, the large
mass neutron stars, and ground-state properties of finite
nuclei~\cite{ji07a,ji07b}. It was known that the associated
parameters that describe the in-medium hadron properties are in
agreement with those from microscopic calculations~\cite{jin95} or
those extracted from recent experimental data at low
densities~\cite{tr05,na06,sch07}. In addition,  the BR scaled NN
interaction was succeeded in yielding the observed $^{14}C$ beta
decay suppression~\cite{ho08}. Therefore, it is appealing to adopt
the models with the BR scaling to study the properties of the LG
phase transition that occurs in the low-density region. The emphasis
is put on the influence of the asymmetric nuclear EOS, especially the
density dependence of the nuclear symmetry energy, on the properties
of the LG phase transition.

\section{Formalism}
\label{rmf}
 The Lagrangian density of the density-dependent RMF models is written as
\begin{eqnarray}\label{equation}
{\mathcal L}&=&\bar\psi\Big[i\gamma_\mu\partial^\mu-M^*+
g_\sigma^*\sigma-g_\omega^*\gamma_\mu\omega^\mu-
g_\rho^*\gamma_\mu\tau_3b_0^\mu\Big]\psi\nonumber\\
&&+\frac{1}{2}\Big(\partial_\mu\sigma\partial^\mu\sigma-
m_\sigma^{*2}\sigma^2\Big)-\frac{1}{4}F_{\mu\nu}F^{\mu\nu}
+\frac{1}{2}m_\omega^{*2}\omega_\mu\omega^\mu\nonumber\\
&& -\frac{1}{4}B_{\mu\nu}B^{\mu\nu}+
\frac{1}{2}m_\rho^{*2}b_{0\mu}b_0^\mu,
\end{eqnarray}
where $\psi,\sigma,\omega$, and $b_0$ are the fields of the nucleon,
scalar, vector, and isovector-vector mesons,  with their in-medium
scaled masses $M^*, m^*_\sigma,m^*_\omega$, and $m^*_\rho$,
respectively. $F_{\mu\nu}$, $ B_{\mu\nu}$ are the strength tensors of
$\om$ and $\rho$ mesons, respectively
\begin{equation}\label{strength} F_{\mu\nu}=\pp_\mu
\om_\nu -\pp_\nu \om_\mu,\hbox{  } B_{\mu\nu}=\pp_\mu b_{0\nu}
-\pp_\nu b_{0\mu},
\end{equation}
$g_\sigma^*$, $g_\omega^*$, and $g_\rho^*$ are the coupling constants
of the scalar, vector and isovector-vector mesons with nucleons,
respectively. The meson coupling constants and hadron masses with
asterisks denote the density dependence, given by the BR
scaling~\cite{ji07a,ji07b,brown91}.

In the mean-field approximation, the energy density and pressure at
the finite temperature are given as
\begin{eqnarray}
\varepsilon
&=&\frac{1}{2}C_\omega^2\rho_B^2+\frac{1}{2}C_\rho^2\rho_B^2\alpha^2+
\frac{1}{2}\tilde{C}_\sigma^2\big(m_N^{*}-M^{*}\big)^2\nonumber\\
&&+\sum_{\tau=p,n}\frac{2}{(2\pi)^3}\int d^3kE^*
[n_\tau(k)+\bar n_\tau(k)],
\end{eqnarray}
and
\begin{eqnarray}
p&=&\frac{1}{2}C_\omega^2\rho_B^2+\frac{1}{2}C_\rho^2\rho_B^2\alpha^2-
\frac{1}{2}\tilde{C}_\sigma^2\big(m_N^{*}-M^{*}\big)^2-\Sigma_0\rho_B\nonumber\\
&&+\frac{1}{3}\sum_{\tau=p,n}\frac{2}{(2\pi)^3}\int d^3k\frac{{\bf
k}^2}{E^*}[n_\tau(k)+\bar n_\tau(k)],\label{pre}
\end{eqnarray}
where $C_\omega = g_\omega^*/m_\omega^*$, $C_\rho =
g_\rho^*/m_\rho^*$,  $\tilde{C_\sigma} = m_\sigma^*/g_\sigma^*$, $E^*
= \sqrt{\vec{k}^2+m_N^*{^2}}$ with $m_N^* = M^*-g_\sigma^*\sigma$ the
effective mass of nucleon, $\alpha$ is the isospin asymmetry
parameter being $\alpha=(\rho_n-\rho_p)/\rho_B$ with
$\rho_B=\rho_n+\rho_p$, and $\Sigma_0$ is the rearrangement term
originating from the density dependence of the parameters,
\begin{eqnarray}
\Sigma_0&=&-\rho_B^2C_\omega\frac{\partial
C_\omega}{\partial\rho_B}-\rho_B^2\alpha^2C_\rho\frac{\partial
C_\rho}{\partial\rho_B}\\\nonumber
  &&-\tilde{C}_\sigma\frac{\partial \tilde{C}_\sigma}{\partial\rho_B}
  \big(m_N^*-M^*\big)^2-\rho_s\frac{\partial M^*}{\partial\rho_B}.
\end{eqnarray}
The distribution functions $n_\tau(k)$ and $\overline n_\tau(k)$ for
nucleon and antinucleon are given as
\begin{eqnarray*}
  n_\tau(k) &=& \{\exp[(E^*(k)-\nu_\tau)/k_BT]+1\}^{-1}, \\
  \overline n_\tau(k) &=& \{\exp[(E^*(k)+\nu_\tau)/k_BT]+1\}^{-1}\hbox{ }(\tau=n,p),
\end{eqnarray*}
where the effective chemical potentials for nucleons read
\begin{equation}
  \nu_\tau = \mu_\tau- g_\omega^*\omega\pm g_\rho^*b_0+\Sigma_0,
\end{equation}
with $+$ in $\pm$ for neutrons and $-$ for protons. The chemical
potentials or effective chemical potentials for nucleons can be
determined from the nucleon densities
\begin{eqnarray}
  \rho_\tau&=&\frac{2}{(2\pi)^3}\int
  d^3k[n_\tau(k)-\overline n_\tau(k)].
  \end{eqnarray}

With the density-dependent RMF models, we can now study the LG phase
transition. In this study, we do not consider the finite size effect
and thus assume as usual the matter is homogeneous.  The two phase
coexistence is governed by the Gibb's conditions:
\begin{eqnarray}
  \mu_\tau(T,\rho_\tau^L) &=& \mu_\tau(T,\rho_\tau^G), \label{gib1}\\
  p(T,\rho_\tau^L) &=& p(T,\rho_\tau^G),\label{gib2}
\end{eqnarray}
where L and G stand for the liquid and gas phase, respectively. The
Gibbs conditions (\ref{gib1}) and (\ref{gib2}) require the same
pressures and chemical potentials for two phases at different
densities and isospin asymmetries.  The stability conditions are
given by
\begin{eqnarray}
\rho_B\Big(\frac{\partial p}{\partial \rho_B}\Big)_{T,\alpha}
&>& 0, \label{bds1} \\
  \Big(\frac{\partial\mu_p}{\partial\alpha}\Big)_{T,p}<0,&
  \mathbf{or} &  \Big(\frac{\partial\mu_n}{\partial\alpha}\Big)_{T,p}>0.\label{bds2}
\end{eqnarray}

\section{Results and discussion}
\label{results}

\begin{figure}[thb]
\begin{center}
\vspace*{-6mm}
\includegraphics[height=10cm,width=10cm]{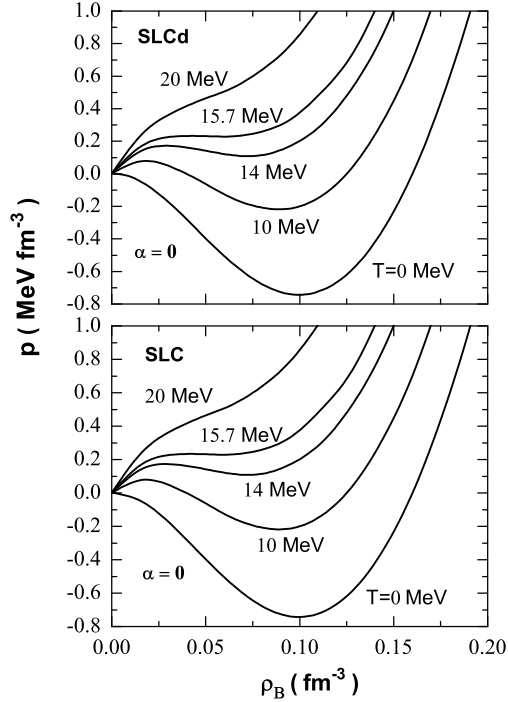}
 \end{center}
\vspace*{-5mm} \caption{The pressure of symmetric nuclear matter
versus the nucleon density $\rho_B$ at different temperatures in
models SLCd  and SLC. \label{prt}}
\end{figure}

We first discuss the liquid-gas phase transition in symmetric nuclear
matter  with the density-dependent RMF models SLC and SLCd that only
differ in the density dependence of the symmetry energy~\cite{ji07b}.
In Fig.~\ref{prt}, we show the pressure of symmetric matter as a
function of nucleon density at various temperatures. At the low
temperature, the pressure first increases and then decreases with the
increasing density. The p-$\rho_B$ isotherms exhibit the form of the
two phase coexistence, with a mechanically unstable  region for each.
In both models that share the same EOS of symmetric
matter~\cite{ji07b}, there appears an equal critical point at
temperature T = 15.7 MeV, where $\partial p/\partial\rho_B =0$, and
$\partial^2p/\partial^2\rho_B =0$. This temperature is called the
critical temperature beyond which symmetric nuclear matter can only
be in a single phase. The experimental value of the critical
temperature extracted for symmetric matter is $16.6\pm0.9$
MeV~\cite{jbn02}, while our prediction is situated on the lower error
bar.

\begin{figure}[thb]
\begin{center}
\vspace*{-4mm}
\includegraphics[height=9cm,width=10cm]{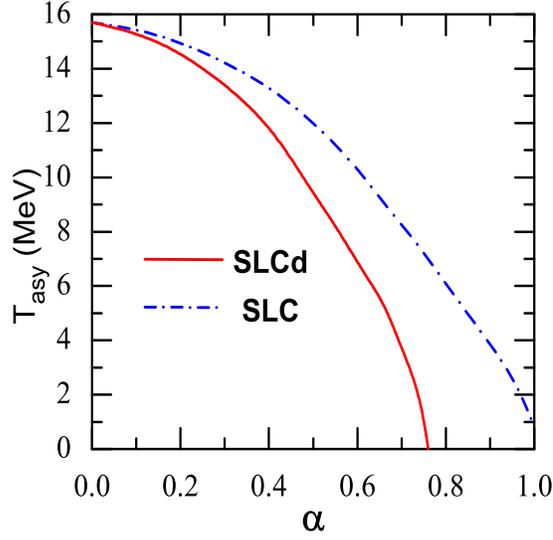}
 \end{center}
\vspace*{-15mm} \caption{(Color online) The temperature $T_{asy}$
versus the asymmetry parameter $\alpha$ in SLCd and SLC. \label{tas}}
\end{figure}

In symmetric matter,  the LG phase transition should be of first
order, whereas in asymmetric matter it was recognized to be of second
order~\cite{hb95,ag12}. In hot asymmetric matter, we can nevertheless
obtain a temperature, denoted as $T_{asy}$, at a fixed $\alpha$
similarly according to the relation $\partial p/\partial\rho_B
=\partial^2p/\partial^2\rho_B =0$. We show in Fig.~\ref{tas} the
$T_{asy}$ as a function of $\alpha$ in SLC and SLCd.   It is found
that this temperature is clearly lower for the softer symmetry energy
that gives higher values in the low-density region where the LG phase
transition occurs. This is understandable because at given
temperatures it is easier to access to a larger isospin asymmetry in
asymmetric matter with a smaller symmetry energy of the model.
However, the temperature determined from a point of inflection is not
the critical temperature for asymmetric matter. In deed, with the
conserved isospin asymmetry of asymmetric matter the pressure and
chemical potentials in the LG coexisting phase change throughout the
transition~\cite{hb95}, in contrast to the case of symmetric matter.
The calculation of the critical temperature $T_c$ in asymmetric
matter is thus different, and we will calculate the $T_c$ later on.

\begin{figure}[thb]
\begin{center}
\vspace*{-4mm}
\includegraphics[height=10cm,width=10cm]{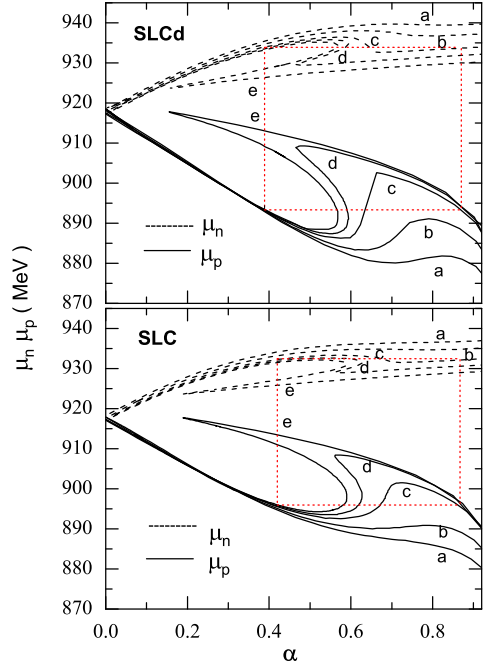}
 \end{center}
\vspace*{-8mm} \caption{(Color online) Chemical potential isobars as
a function of $\alpha$ at temperature T = 10 MeV. The curves a, b, c,
d,  and e correspond to the pressures p = 0.235, 0.175, 0.12, 0.098,
and 0.08 $MeV fm^{-3}$ in SLCd and SLC, respectively. The rectangle,
an example for the geometrical construction for $p = 0.12 MeV
fm^{-3}$, is used to determine the asymmetry parameters and chemical
potentials in the LG coexistence phase.
 \label{chal}}
\end{figure}

Now, we turn to the discussion on the LG phase transition  in
asymmetric nuclear matter according to the Gibbs conditions
(\ref{bds1}) and(\ref{bds2}).  For a fixed pressure, the two
solutions of the LG phases to the Gibbs conditions form the edges of
a rectangle in the proton and neutron chemical potential isobars as a
function of isospin asymmetry $\alpha$ and can be found by means of
the geometrical construction method~\cite{hb95,qwl00}. We thus need
firstly to obtain the proton and neutron chemical potential isobars.
The chemical potentials of the proton and neutron, together with the
isospin asymmetry parameter, are numerically determined at given
pressures and densities. Fig.~\ref{chal} shows the $\mu_n$ and
$\mu_p$ isobars as a function of $\alpha$  at fixed temperature $T =
10$ MeV for various pressures. The solid and dashed curves are for
protons and neutrons, respectively. We see that the curves for lower
pressures are more complicated than those for higher pressures. Using
the geometrical construction in the chemical potential isobars, we
can then obtain two solutions at different isospin asymmetry
parameters $\alpha$ for two phases under the chemical and mechanical
equilibriums. As an example, we plot the rectangle for the case of $T
= 10 MeV$ and $p = 0.12 MeV fm^{-3}$ in Fig.~\ref{chal}. The solution
with the larger $\alpha$ defines the gas phase with the lower
density, while the solution with the smaller $\alpha$ defines the
liquid phase with the higher density.

\begin{figure}[thb]
\begin{center}
\vspace*{-6mm}
\includegraphics[height=8.5cm,width=9.5cm]{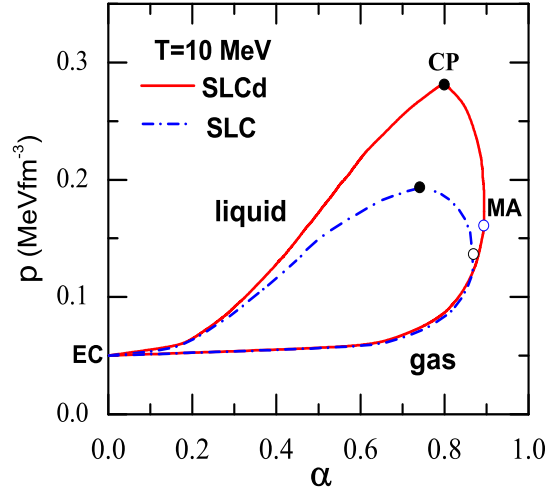}
 \end{center}
\vspace*{-15mm} \caption{(Color online) The  section of the binodal
surface at T = 10 MeV in SLCd and SLC.  The CP, MA and EC stand for
the critical point, the maximal isospin asymmetry and the point of
equal concentration, respectively. \label{pa}}
\end{figure}

\begin{figure}[thb]
\begin{center}
\vspace*{-4mm}
\includegraphics[height=8.5cm,width=9.5cm]{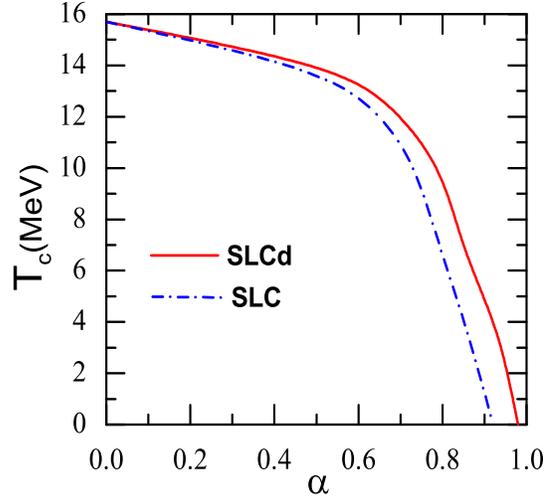}
 \end{center}
\vspace*{-15mm} \caption{(Color online) The critical temperature
$T_{c}$ versus the asymmetry parameter $\alpha$ in SLCd and SLC.
\label{ta}}
\end{figure}

The pair of the solutions of the Gibbs conditions found by the
geometrical method forms the phase-separation boundary at a given
pressure in hot asymmetric nuclear matter, while all pairs build up
the binodal surface. In Fig.~\ref{pa}, we show the section of the
binodal surface at T=10 MeV. The binodal surface is divided into two
branches by the critical point (CP) ($\alpha_c$, $p_c$) and the point
of the equal concentration (EC) ($\alpha=0$). One branch is the
high-density (liquid) phase, and the other branch is the low-density
(gas) phase. We see in Fig.~\ref{pa} that a much higher CP and a much
larger boundary of the phase coexistence appear in SLCd, compared
with those in SLC. This indicates that the section of the binodal
surface is very sensitive to the density dependence of the symmetry
energy, because the unique difference between the SLCd and SLC is
that the former possesses  the much softer symmetry energy than that
in the latter. Similar sensitivity was found in
Refs.~\cite{jx07,bk09}. Also, we see that the softness of the
symmetry energy in SLCd makes the maximal asymmetry (MA) a little
larger, as compared with the MA in SLC.

\begin{figure}[thb]
\begin{center}
\vspace*{-4mm}
\includegraphics[height=10cm,width=12cm]{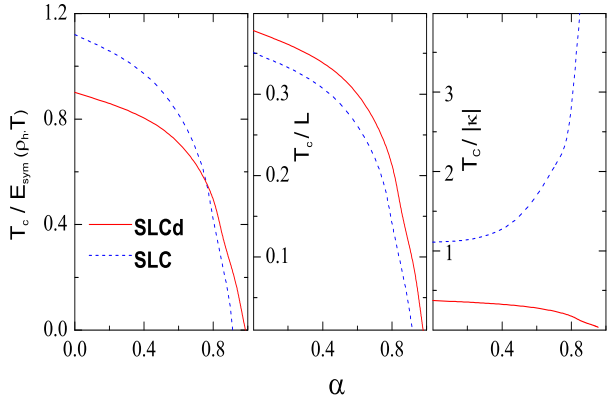}
 \end{center}
\vspace*{-40mm} \caption{(Color online) Ratios of the critical
temperature $T_{c}$ to the symmetry energy (left), the slope $L$
(middle) and the curvature modulus $|\kappa|$ (right) as a function
of the isospin asymmetry. The slope  and curvature of the symmetry
energy are also calculated at density $\rho_h$. \label{tsym}}
\end{figure}

As shown in the binodal curve in Fig.~\ref{pa}, the system can only
be in a single phase for $\alpha>\alpha_c$ at a given temperature.
Correspondingly,  at a fixed $\alpha$ ($\alpha=\alpha_c$) the given
temperature plays a role of the critical temperature $T_{c}$ beyond
which the system totally cannot fall into the liquid phase at all
pressures. In this way, one can obtain the critical temperatures at
various $\alpha$. In Fig.~\ref{ta}, we depict $T_{c}$ as a function
of $\alpha$ in SLC and SLCd. It is seen that the critical temperature
separating a single phase from the coexisting liquid-gas phase
decreases rapidly with increasing the isospin asymmetry parameter
$\alpha$ $\geq$ 0.6. We find from Fig.~\ref{ta} that $T_{c}$ is
higher for the model with the softer symmetry energy (SLCd) that
gives higher value of the symmetry energy in the low-density region
where the LG phase transition occurs. Our finding is in agreement
with that in Ref.~\cite{bk09}. Interestingly, we find that the
sensitivity of $T_c$ to differences in the symmetry energy is exactly
in contrast to that of $T_{asy}$ as shown in Fig.~\ref{tas}. As seen
in Fig.~\ref{ta},  $T_c$ is not very typically sensitive to
differences in the symmetry energy. In deed, the symmetry energy
relies also on the temperature. In order to exhibit the dependence of
the critical temperature on the symmetry energy, we draw in
Fig.~\ref{tsym} the ratios of the critical temperature $T_{c}$ to the
symmetry energy, the slope and the curvature modulus as a function of
the isospin asymmetry. The symmetry energy is here expanded at
density $\rho_h=0.08fm^{-3}$ as
\begin{equation}\label{eqsym1}
    E_{sym}(\rho_B,T)=E_{sym}(\rho_h,T)+\frac{L}{3}\chi_h
+\frac{\kappa_{sym}}{18}\chi_h^2+\cdots,
\end{equation}
where $L$ and $\kappa_{sym}$ are the slope and curvature of the
symmetry energy at $\rho_h=0.08fm^{-3}$, and
$\chi_h=(\rho_B-\rho_h)/\rho_h$. The reason for choosing
$\rho_h=0.08fm^{-3}$ is due to the facts that the critical density at
the critical point is below but close to 0.08$fm^{-3}$ and that this
density may be regarded as an average density over the surface and
volume of finite nuclei~\cite{chen11}. We see from Fig.~\ref{tsym}
that the ratios of the $T_c$ to the symmetry energy and its slope are
just moderately different for the SLCd and SLC, while the distinction
is surprisingly large for the curvature. This demonstrates that at
the given symmetry energy and slope the $T_c$ is moderately different
for the SLCd and SLC, consistent with the result shown in
Fig.~\ref{ta}. Though in the right panel of Fig.~\ref{tsym} the large
departure can be observed for the $T_c$ at given curvatures, the
final role of the curvature is restrained by the factor $\chi_h^2$
which is quite small in the present case. Moreover, we find that the
slope and curvature terms in Eq.(\ref{eqsym1}) play an opposite role
to that of the zero-order term in affecting the critical temperature.
As a result, the critical temperature in hot asymmetric matter is
just moderately model dependent, as shown in Fig.~\ref{ta}.

At last, it is interesting to point out that the limiting pressure in
the binodal surface does not appear in the present models. As pointed
out in Refs.~\cite{qwl00,cw11}, the limiting pressure that is a
cutoff of the pressure beyond which the pair of the solution does not
exist in the geometrical construction is a result of the density
dependence of the $\rho$ meson-nucleon vertex. However, this seems to
be rather model-dependent. In deed, the difference in values of the
limiting pressure is large in models applied in
Refs.~\cite{qwl00,cw11}. In some nonlinear RMF models~\cite{bk09}
where the density dependence of the $\rho$ meson-nucleon interaction
is actually induced by  $\rho$-meson effective mass, the limiting
pressure also does not appear.

\section{SUMMARY}
\label{summary}

In summary, we have investigated the effects of the nuclear EOS with
the density-dependent interactions on the liquid-gas phase transition
in symmetric and asymmetric nuclear matter with the well-constrained
RMF models SLCd and SLC that feature in-medium hadron properties
according to the BR scaling and can provide satisfactory description
for properties of finite nuclei and neutron stars. The critical
temperature of the LG phase transition in symmetric matter is
obtained to be 15.7 MeV for both models that have the same symmetric
part of the nuclear EOS, in nice agreement with experimental data.
With an analytic continuation from symmetric to asymmetric matter, we
have obtained a temperature $T_{asy}$ in hot asymmetric matter
according to the monotonicity of the pressure and demonstrated that
this temperature is lower for the softer symmetry energy.  It is
found in hot asymmetric matter that the boundary of the
phase-coexistence region is very sensitive to differences in the
symmetry energy in SLC and SLCd, while the dependence on the EOS of
symmetric matter is much weaker. A higher pressure and a larger area
of the LG phase coexistence region appear with a softer symmetry
energy. The critical temperature of hot asymmetric matter that
separates the single-phase region from the two-phase region is found
to be higher for the model with the softer symmetry energy, while the
magnitude of the sensitivity is analyzed to be just moderate as
compared with that for the boundary of the LG phase coexistence.

\section*{Acknowledgement}

We would like to thank Lie-Wen Chen and Chen Wu for useful
discussions. The work was supported in part by the National Natural
Science Foundation of China under Grant Nos. 10975033 and 11275048
and the China Jiangsu Provincial Natural Science Foundation under
Grant No.BK2009261.

\end{document}